# Unusual nodal behaviors of the superconducting gap in the iron-based superconductor Ba(Fe$_{0.65}$Ru$_{0.35}$)$_2$As$_2$: Effects of spin-orbit coupling

L. Liu[1], K. Okazaki[1,2], T. Yoshida[3], H. Suzuki[1], M. Horio[1], L. C. C. Ambolode II[1], J. Xu[1], S. Ideta[1], M. Hashimoto[4], D. H. Lu[4], Z.-X. Shen[4], Y. Ota[2], S. Shin[2], M. Nakajima[5], S. Ishida[5], K. Kihou[6], C. H. Lee[6], A. Iyo[5], H. Eisaki[5], T. Mikami[1], T. Kakeshita[1], Y. Yamakawa[7], H. Kontani[7], S. Uchida[1], and A. Fujimori[1]

[1]Department of Physics, University of Tokyo, Bunkyo-ku, Tokyo 113-0033, Japan
[2]Institute for Solid State Physics (ISSP), University of Tokyo, Kashiwa, Chiba 277-8581, Japan
[3]Graduate School of Human and Environmental Studies, Kyoto University, Kyoto 606-8502, Japan
[4]Stanford Synchrotron Radiation Lightsource, SLAC National Accelerator Laboratory, Menlo Park, California 94025, USA
[5]Electronics and Photonics Research Institute, National Institute of Advanced Industrial Science and Technology, Tsukuba 305-8568, Japan
[6]Research Institute for Energy Conservation, National Institute of Advanced Industrial Science and Technology, Tsukuba 305-8568, Japan
[7]Department of Physics, Nagoya University, Furo-cho, Nagoya 464-8602, Japan

We have investigated the superconducting gap of optimally doped Ba(Fe$_{0.65}$Ru$_{0.35}$)$_2$As$_2$ by angle-resolved photoemission spectroscopy (APRES) using bulk-sensitive 7 eV laser and synchrotron radiation. It was found that the gap is isotropic in the $k_x$-$k_y$ plane both on the electron and hole Fermi surfaces (FSs). The gap magnitudes of two resolved hole FSs show similar $k_z$ dependences and decrease as $k_z$ approaches $\sim 2\pi/c$ (i.e., around the Z point) unlike the other Fe-based superconductors reported so far, where the superconducting gap of only one hole FS shows a strong $k_z$ dependence. This unique gap structure can be understood in the scenario that the $d_{z^2}$ orbital character is mixed into both hole FSs due to the finite spin-orbit coupling between almost degenerate FSs and is reproduced by calculations within the random phase approximation including the spin-orbit coupling.

The momentum dependence of the superconducting (SC) order parameter in the Fe-based superconductors (FeSCs) has been intensively investigated by angle-resolved photoemission spectroscopy (ARPES) due to their close relationship with paring mechanisms. A variety of superconducting gap symmetries in FeSCs, such as $s_\pm$-wave, nodeless $s_{++}$ and nodal $s$-wave, have been discussed from the theoretical point of view [1, 2]. Up to now, even for the most intensively studied 122 systems, the situation is still complicated. As for the isovalently substituted BaFe$_2$(As$_{1-x}$P$_x$)$_2$ compound, it was revealed by the penetration depth and thermal conductivity measurements that there exist line nodes in the SC gap (for $x = 0.33$) [3]. It has been pointed out based on the random phase approximation (RPA) calculation that, three-dimensional (3D) horizontal line nodes may appear around the Z point [$k = (0, 0, 2\pi/c)$] on the outermost strongly warped hole Fermi surface (FS) which has dominantly $d_{z^2}$ orbital character [4]. However, ARPES experiments have not come to agreement on the existence of the horizontal line nodes on the 3D hole FS of BaFe$_2$(As$_{0.7}$P$_{0.3}$)$_2$ [5, 6].

Ba(Fe$_{1-x}$Ru$_x$)$_2$As$_2$, another isovalently substituted system, in which doping is done at the pivotal Fe site [7, 8], also exhibits a highly warped hole FS along the $k_z$ direction [9, 10]. Recent thermal conductivity measurements suggest that Ba(Fe$_{1-x}$Ru$_x$)$_2$As$_2$ has also SC gap line nodes for a wide doping range [11]. In order to clarify the SC gap structure and the location of the suggested line nodes in momentum space, we performed an ARPES study of optimally doped Ba(Fe$_{0.65}$Ru$_{0.35}$)$_2$As$_2$.

In this study, two kinds of APRES apparatus were employed: one is bulk-sensitive ARPES using a vacuum ultraviolet laser; the other is based on synchrotron radiation, which allows us to trace the $k_z$ dependence of the SC gap by tuning the incident photon energy. We show that the SC gaps both on the hole FSs and electron FSs are isotropic in the $k_x$-$k_y$ plane, while the SC gaps on the hole FSs are strongly $k_z$ dependent. Interestingly, both of the two resolved hole FSs exhibit SC gaps with almost the same $k_z$ dependence and magnitudes. We attribute this unique observation to the effect of orbital fluctuations as well as of the spin-orbit coupling which arises from the Fe 3d, Ru 4d, and As 4p electrons.

Single crystals of Ba(Fe$_{0.65}$Ru$_{0.35}$)$_2$As$_2$ were grown by the self-flux method and post-annealed as described in Refs. [12, 13]. Magnetization measurements and resistivity measurements indicated that superconductivity appeared below $\sim 21$ K, as shown in Fig. S1 [14]. The superconducting transition width was 2-3 K, broader than that of BaFe$_2$(As$_{1-x}$P$_x$)$_2$ which shows a typical width of less than 0.5 K [15]. This is presumably due to the inhomogeneity of Ru content in the single crystals, arising from the larger ion size of Ru than Fe.

Laser-ARPES experiments were carried out at the Institute of Solid State Physics (ISSP, Japan) [16]. A VG-Scienta HR8000 analyzer and a 6.994 eV quasi-CW laser were used with the total energy resolution ($\Delta E$) set to $\sim 3$ meV for the FS mapping and $\sim 1.2$ meV for the SC gap measurement. The angular resolution was 0.1°. Crystals were cleaved in-situ at $T = 3$ K in an ultrahigh vacuum better than $2 \times 10^{-11}$ Torr. ARPES experiments using synchrotron light were performed at beamline 5-4 of Stanford Synchrotron Radiation Lightsource (SSRL) with a Scienta R4000 analyzer. The energy resolution was ~4-8 meV for the gap measurement depending on the photon energy. Crystals were cleaved in-situ at 6 K in an ultrahigh vacuum of ~2.5×10$^{-11}$ Torr. Calibration of the Fermi level ($E_F$) of the samples was achieved by referring to that of gold. In-plane ($k_x$, $k_y$) and out-of-plane momenta ($k_z$) are expressed in units of $\pi/a$ and $2\pi/c$, respectively, where $a$ = 4.03 Å and $c$ = 12.76 Å are the in-plane and out-of-plane lattice constants [8, 17]. Here, $x$ and $y$ axes point toward the tetragonal (100) direction.

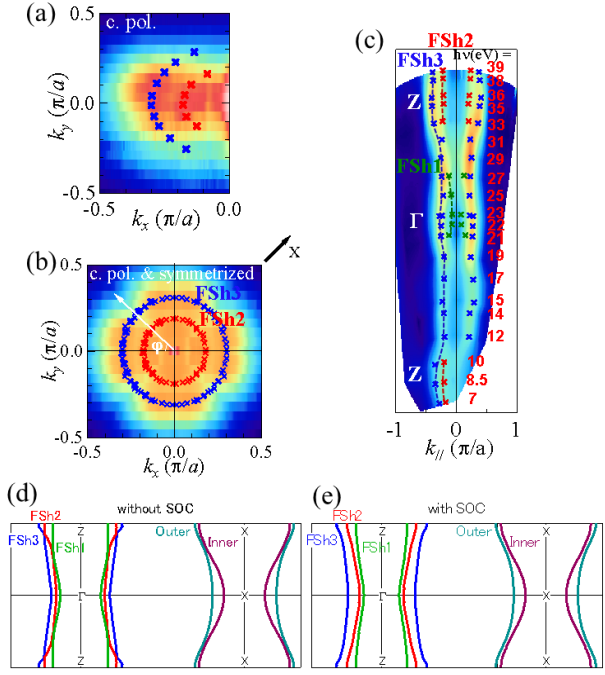

FIG. 1. (a) FS intensity map of Ba(Fe$_{0.65}$Ru$_{0.35}$)$_2$As$_2$ measured at 25 K probed by circularly polarized 7 eV photons. The integration energy window is ± 5 meV around $E_F$. (b) FS image map obtained by symmetrizing the map of panel (a), taking into account the tetragonal fourfold rotational symmetry. The red and blue markers indicate the positions of $k_F$'s for the two hole FSs, namely FSh2 and FSh3, respectively. Definition of the FS angle φ is also shown. (c) Photon energy dependence of the ARRES intensity map measured at 6 K, demonstrating the $k_z$ dependence of hole FSs. The linearly polarized light was used and the measured cut crosses the BZ center along the $x$ ($y$) axis. Each FS is labeled from the comparison to the calculated FSs. (d), (e) Calculated FSs of BaFe$_2$As$_2$ based on DFT with and without SOC. The pnictogen height from the iron layer $h_{pn}$ is reduced by 2.5 % from the reported value for BaFe$_2$As$_2$ so as to make the structural parameters close to that of Ba(Fe$_{0.65}$Ru$_{0.35}$)$_2$As$_2$.

In order to investigate the effects of spin-orbit coupling (SOC) on the orbital character of Fermi surfaces, we have performed band-structure calculations based on density-functional theory (DFT) with and without SOC for BaFe$_2$As$_2$ and BaRu$_2$As$_2$ using a WIEN2k package [19]. In order to investigate the electronic structure of Ba(Fe$_{0.65}$Ru$_{0.35}$)$_2$As$_2$, we use the lattice parameters of BaFe$_2$As$_2$ ($a$ = 3.96 Å and $c$ = 13.02 Å) and the pnictogen height $h_{pn}$ reduced by 2.5 % from that of BaFe$_2$As$_2$ ($h_{pn}$ = 1.326 Å) for the calculations of both BaFe$_2$As$_2$ and BaRu$_2$As$_2$. SOC is included in the Fe 3d and Ru 4d orbitals only.

We also have performed RPA calculations to investigate the $k_z$ dependence of the SC gap for BaFe$_2$As$_2$ with the structure described above also with and without SOC. The details for calculations are described in Ref. [2] for RPA and in Ref. [20] for the inclusion of SOC in SC-gap calculation.

Figure 1 shows the FS intensity map around the Brillouin zone (BZ) center. Two hole FSs can be resolved using 7 eV photons as indicated by red and blue symbols, respectively, in Fig. 1(a). In order to have a full view of the hole FSs, we have symmetrized the experimentally obtained FS intensity map and obtained the image map of Fig. 1(b). Both of the two resolved FSs are almost circular. Figure 1(c) presents $k_z$ mapping of the hole FSs by tuning the photon energy using synchrotron radiation. Assuming that the inner potential is 14.5 eV, the corresponding $k_z$ fitted under the free-electron final state model is $k_z$ = 0 at ~23 eV and $2\pi/c$ at ~35 eV. On the other hand, in the region of low photon energy (< 21 eV), relatively large FSs appear around 8.5 eV, implying the location of another Z point, while the free-electron final state model appears to become inaccurate. Based on the $k_z$ mapping, 7 eV photons used in the laser-ARPES experiments probe the hole FS profile near the Z point. Here, two hole FSs can be resolved throughout one period of the momentum space along the $k_z$ direction, while the band-structure calculation for Ba(Fe$_{0.62}$Ru$_{0.38}$)$_2$As$_2$ found three hole FSs around the Z point. We could not identify another hole FS around the Z point in the ARPES spectra, which is probably attributed to the weak spectral intensity, e. g., of the $d_{xy}$ band [18]. We have compared the $k_z$ mapping with the results of the DFT calculations with and without SOC as shown in Figs. 1(d) and 1(e), respectively, and assigned the two resolved FSs to the FSh1 and FSh3 surfaces near the Γ point and to FSh2 and FSh3 near the Z point, because the $d_{xy}$ orbital, which shows only very weak intensity, contributes to FSh2 near the Γ point and to FSh1 near the Z point according to the DFT calculations.

Figure 2 presents the SC gap probed by 7 eV laser photons. The temperature evolution of the energy distribution curves (EDCs) at a typical $k_F$ point on the FSh3 are shown in Fig. 2(e). With decreasing temperature below $T_c$, the leading edge shifts to the lower energy and a peak-like structure develops, clearly indicating the opening of a SC gap. The FS angle φ is defined as indicated in Fig. 1(b). Figures 2(a)-2(d) show the EDCs at the $k_F$ points below $T_c$ (~3 K) and above $T_c$ (~25 K) for various FS angles. In order to remove the effect of the Femi-Dirac cutoff, EDCs were symmetrized with respect to $E_F$. One can see a significant loss of intensity around $E_F$ and the formation of peak-like structures below and above $E_F$ in the symmetrized EDCs. To reveal the SC gap energy, vertical solid bars are shown where the intensity begins to weaken in the symmetrized EDCs. The observed SC gap on the hole FSs of Ba(Fe$_{0.65}$Ru$_{0.35}$)$_2$As$_2$ is almost isotropic and independent of FSs with the magnitude of $\Delta \sim 2.5$ meV.

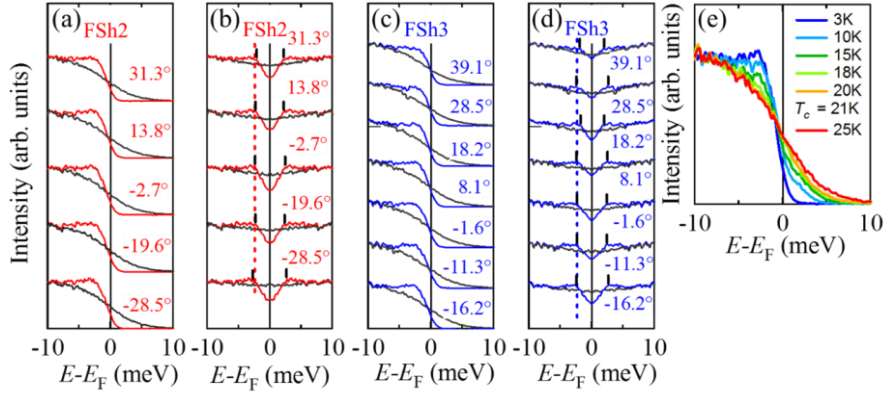

FIG. 2. EDCs at the $k_F$ points probed by 7 eV laser photons. (a-d) EDCs and symmetrized EDCs for various FS angles. The red(blue) lines show the spectra on the FSh2 (FSh3) below $T_c$ (~3 K) and the black ones are above $T_c$ (~25 K). Vertical solid bars indicate the gap energy. The dashed lines are guides to the eye at the energy of 2.5 meV. (e) Typical temperature dependence of EDCs on the FSh3 at $\varphi = -16°$.

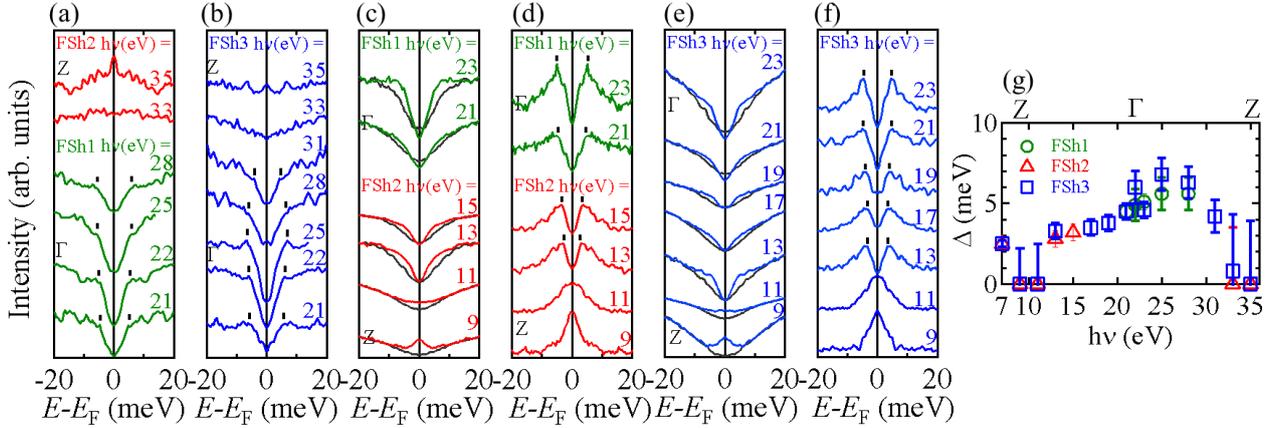

FIG. 3. EDCs at the $k_F$ points with $\varphi = 0°$ probed by synchrotron radiation using various photon energies and hence at various $k_z$'s. (a-b) Symmetrized EDCs on the two resolved FSs below $T_c$ ($T \sim 6$ K) using linearly polarized 21-35 eV photons. (c) Symmetrized EDCs on the inner one of the two resolved FSs (FSh1 or FSh2) taken with circularly polarized 9-23 eV photons. Red curves show the spectra below $T_c$ ($T \sim 6$ K) and black ones above $T_c$ ($T \sim 25$ K). (d) Spectra obtained by dividing the symmetrized EDCs below $T_c$ by those above $T_c$ in panel (c). (e-f) The same spectra on the FSh3 as panels (c) and (d), respectively. (g) SC gap size ($\Delta$) plotted as a function of incident photon energy, showing a strong $k_z$ dependence.

Next, we studied the $k_z$ dependence of the SC gap on the hole FSs by tuning the incident photon energy. In Figs. 3(a)-3(b), one can see a clear gap both on the two resolved FSs in the symmetrized EDCs probed by 22 eV photons, whereas no clear gap opening could be observed in the 33 eV and 35 eV spectra. Since the energy resolution of the measurements using high energy photons may not be enough to detect small SC gaps, low energy photons (hν = 9-23 eV) with relatively high energy resolution ($\Delta E$ = 4.5-6 meV) were used to investigate the $k_z$ dependence, as shown in Figs. 3(c)-3(f). In order to clarify the gap magnitude, we have divided the symmetrized EDCs at the $k_F$ positions below $T_c$ by that above $T_c$ and obtained the spectra shown in Figs. 3(d) and 3(f). Figure 3(g) summarizes the $k_z$ dependence of the SC gap magnitude, where the data obtained by 7 eV laser-ARPES is also plotted. The SC gap shows the largest magnitude of ~6 meV around the Γ point. When $k_z$ moved from the Γ to Z points, the gap size became small and a clear gap anisotropy along the $k_z$ direction was observed. The gap magnitude obtained by laser-ARPES having the energy resolution as high as ~1.2 meV, is consistent with the $k_z$ dependence of the measured SC gap using synchrotron radiation. One can also see that no gap was clearly observed by 9 eV, 11 eV, 33 eV and 35 eV photons. Note that observing the split peaks below and above $E_F$ in the symmetrized EDCs with the SC gap opening is subject to the energy resolution $\Delta E$. We achieved a high energy resolution of $\Delta E$ = 4-5 meV using 9 eV and 11 eV photons. Although in both cases two peaks merge into one peak located at $E_F$, Figs. 3(d) and 3(f) present that the 11 eV spectra are broader than the 9 eV ones for both the FSh2 and FSh3, implying that tiny gap opening in the 11 eV spectra might be masked due to the limited resolution. Although it is difficult to conclude whether the gap magnitude probed by 9 eV photons becomes zero or a small non-zero value, at least it becomes smaller as $k_z$ approaches the Z point. This suggests that the gap minimum appears around the Z point. Furthermore, our measurements show that the two resolved hole FSs show almost equivalent gap size and show similar momentum dependence along the $k_z$ direction.

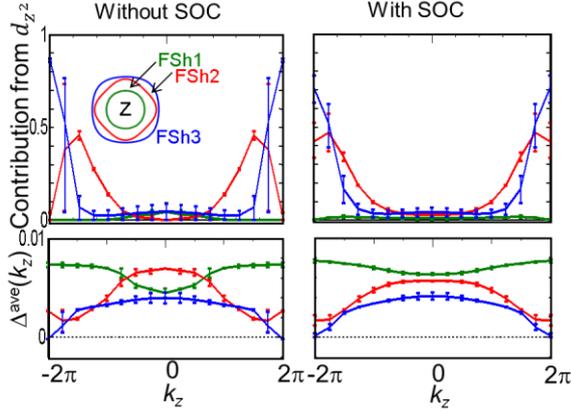

FIG. 4. $k_z$ dependence of the SC gap size Δ calculated by RPA with and without SOC. The same crystal structure as that for the DFT calculations shown in Fig. 1 was used.

Saito *et al.* calculated the momentum dependences of SC gaps based on a model which takes into account strong spin and orbital fluctuations. Irrespective of the relative strengths of the spin and orbital fluctuations, the SC gap magnitude shows a minimum for $k_z$ around the Z point only if the FS has strong $d_{z^2}$ orbital character [2]. Based on the calculation, therefore, it is difficult to understand the present APRES observations that both two resolved hole FSs exhibit a gap minimum around the Z point.

In the previous band-structure calculation it was found that the $d_{z^2}$ orbital mainly contribute to the outermost FS with the strongest $k_z$ dispersion around the Z point. Since most of previous studies of band dispersions of FeSCs did not include SOC in the calculations, we have performed DFT band-structure calculations including SOC for BaFe$_2$As$_2$ with the reduced pnictogen height corresponding to the crystal structure of Ba(Fe$_{0.62}$Ru$_{0.38}$)$_2$As$_2$ using a WIEN2k package [19], and results are shown in Fig. S2 [14]. For the calculation without SOC, we found that the $d_{z^2}$ orbital around the Z point mainly contribute to one of the three hole FSs, in good agreement with previous theoretical results for BaFe$_2$As$_2$ [10]. Compared to the results for BaFe$_2$As$_2$, the top of the $d_{z^2}$ band is shifted upward at the Z point and the FSs of the $d_{z^2}$ and $d_{yz/zx}$ bands become almost degenerate. By including SOC, $d_{z^2}$ orbital is mixed to the FS that is mainly contributed from the $d_{yz/zx}$ orbital in the calculation without SOC. We have also performed calculations for BaRu$_2$As$_2$ with the same structure, and the results shown in Fig. S3 [14]. Compared to the results for BaFe$_2$As$_2$, the band widths are broadened nearly twice and the admixture of the $d_{z^2}$ orbital is comparable to BaFe$_2$As$_2$. These findings demonstrate an important modification of the orbital character of FSs due to SOC when almost degenerate FSs exist. This implies that the effect of SOC on modifying the orbital character depends on details of the band structure. Under such circumstances, not only $d_{z^2}$ intra-band but also inter-band scattering tends to occur, resulting in the observation that the gap magnitudes both on the two resolved hole FSs varies similarly with $k_z$. We also note that the SOC-induced modification of the FS topology and its consequent effects on the SC gap structure have been recently discussed in LiFeAs based on the orbital-spin fluctuation theory [20]. These results suggest important roles of SOC in a variety of SC gap structures of FeSCs.

In order to confirm the above scenario, we have performed RPA calculations to investigate the $k_z$ dependence of the SC gap for BaFe$_2$As$_2$ with the above described crystal structure with the reduced pnictogen height with and without SOC, and the results are shown in Fig. 4. Error bars correspond to the in-plane anisotropy at a specific $k_z$. For the calculation without SOC, the $k_z$ dependence of the SC gap is similar to the previous results for BaFe$_2$(As,P)$_2$ [2], that is, the $k_z$ dependence of the SC gap is strong only for the FS dominantly contributed from the $d_{z^2}$ orbital. In the calculation including SOC, contributions from the $d_{z^2}$ orbital become comparable between FSh2 and FSh3, and the SC gaps of these two FSs are $k_z$-dependent and minimized around $k_z = 2\pi$, which seems consistent with the experimental results.

Finally, we comment on the existence of line nodes in the SC gap of this compound suggested by thermal conductivity measurements, in which large residual thermal conductivity at zero field and its dependence on $\sqrt{H}$ have been found [11]. "Vertical" line nodes have been revealed in heavily K-doped 122 materials [16, 21], while our ARPES observations of the isotropic SC gap opening in the $k_x$-$k_y$ plane rule out this possibility. Although it is difficult to make a definite conclusion due to the intrinsic $k_z$ broadening and the limited energy resolution of ARPES experiment, considering the $k_z$ dependence of the gap observed in this compound, it is likely that horizontal gap nodes, if existing, appear around the Z point.

Our ARPES studies on Ba(Fe$_{0.65}$Ru$_{0.35}$)$_2$As$_2$ demonstrated an isotropic superconducting gap structure in the $k_x$-$k_y$ plane both on the hole and electron FSs. By tuning the photon energy of synchrotron radiation, a clear momentum dependence of the SC gap on the hole FSs along the $k_z$ direction was revealed, consistent with the laser-APRES measurements with a fixed photon energy. Moreover, we found that the gap size on both two resolved hole FSs varied similarly with $k_z$ and exhibits a gap minimum around the Z point. From band-structure calculation including SOC, we found that the $d_{z^2}$ orbital character is mixed in both FSs around the Z point due to SOC and that the RPA calculation including SOC successfully reproduce the observed gap structure. These findings have profound implications for the important role of SOC in understanding the variety of gap structures found for FeSCs.

We are grateful to D. Hirai for his help in the magnetic susceptibility measurement. This work was supported by a Grant-in-Aid for Scientific Research from the Japan Society for the Promotion of Science (JSPS). The

Stanford Synchrotron Radiation Lightsource is operated by the Office of Basic Energy Science, US Department of Energy. Part of this work was conducted in Research Hub for Advanced Nano Characterization, University of Tokyo, supported by the Ministry of Education, Culture, Sports, Science and Technology (MEXT), Japan. L.L. thanks the MEXT Scholarship Program of Japan and the China Scholarship Council (CSC) for financial support. H.S. and M.H. acknowledge financial support from the Advanced Leading Graduate Course for Photon Science (ALPS) and the JSPS Research Fellowship for Young Scientists.

**Supplemental Material**

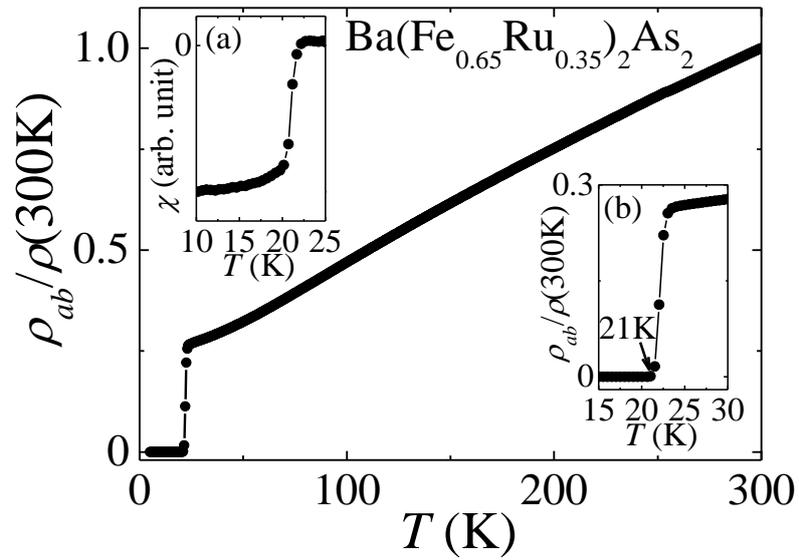

FIG. S1. Temperature dependence of the in-plane resistivity of Ba(Fe$_{0.65}$Ru$_{0.35}$)$_2$As$_2$. Inset (a) presents the temperature dependence of the dc magnetic susceptibility. Inset (b) shows an enlarged plot of the resistivity near $T_c$.

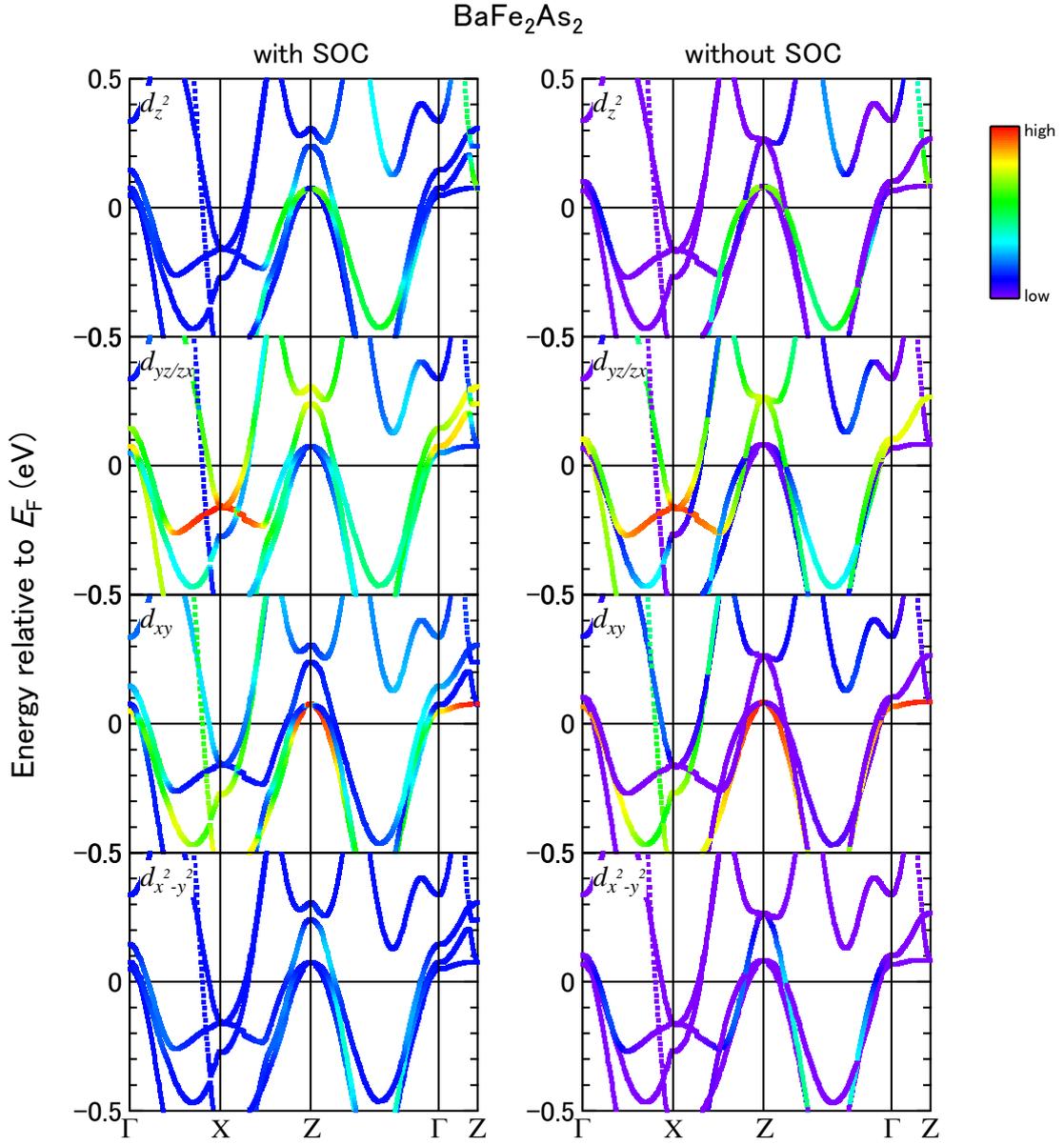

FIG. S2. DFT band dispersions of BaFe$_2$As$_2$ near $E_F$ calculated with and without SOC. The crystal structure parameters used for these calculations are the same as those used in Figs. 1(d) and 1(e).

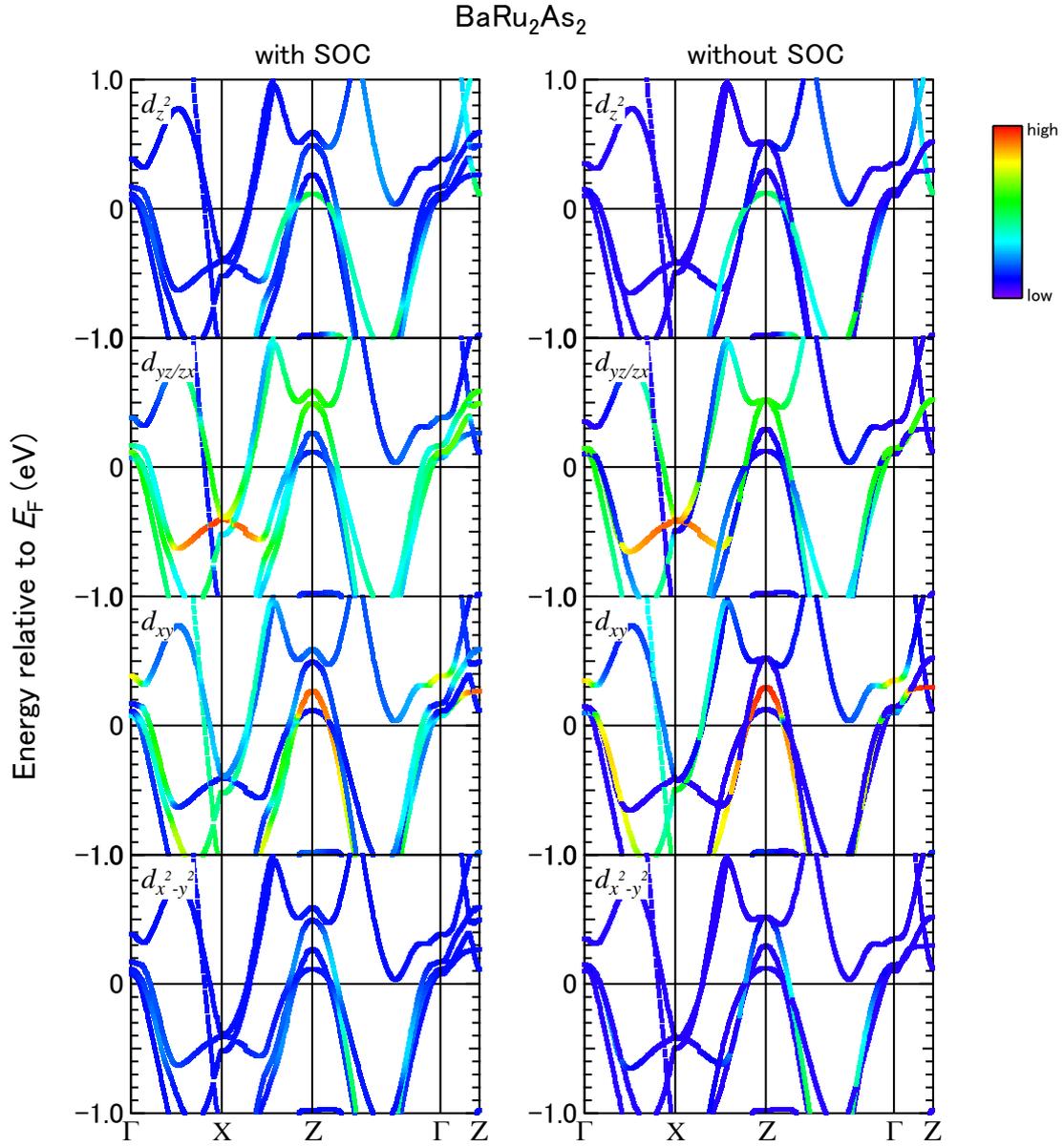

FIG. S3. DFT band dispersions of BaRu$_2$As$_2$ near $E_F$ calculated with and without SOC. The crystal structure parameters used for these calculations are the same as those used for BaFe$_2$As$_2$, for which results are shown in Figs. 1(d), 1(e), and S2. Compared to the results shown in Fig. S2, the band widths are generally broadened nearly twice and the admixture of the $d_{z^2}$ orbital is comparable to BaFe$_2$As$_2$. Note that the plotted energy range is from -1.0 to 1.0 eV for this figure, whereas that for Fig. S2 is from -0.5 to 0.5 eV.

FigureS4 presents the SC gap on the electron FSs probed by 21 eV photons. Two electron FSs can be clearly resolved as shown in Figs. S4(a) and S4(b). Supposing that the inner potential of 14.5 eV is still applicable, $k_z$ around the zone corner probed by 21 eV photons is ~$5.3(2\pi/c)$. For this $k_z$ value, the observed FS shapes are found to be consistent with the calculated shapes of the electron FSs. Symmetrized EDCs at various $k_F$ points measured below $T_c$ are presented in Figs. S4(c) and S4(d), showing an almost isotropic SC gap magnitude in the $k_x$-$k_y$ plane both on the outer and inner electron FSs.

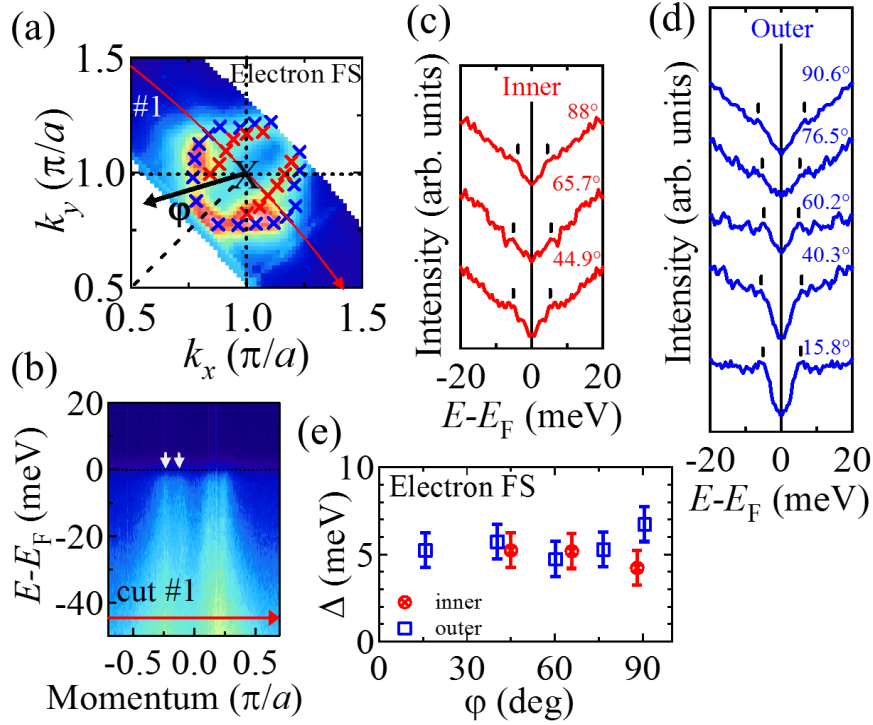

FIG. S4. Electron FSs, band dispersions, and SC gaps on electron FSs. (a) Electron FS mapping around the zone corner taken with linearly polarized 21 eV photons. The definition of the FS angle φ on the electron FSs is indicated. (b) Energy-momentum plot corresponding to cut #1 in panel (a). (c-d) Symmetrized EDCs at the $k_F$ points of outer FS and inner FS measured below $T_c$ ($T \sim 6$ K) for various FS angles. Vertical bars indicate the SC gap energy. (e) SC gap size in panels (c-d) plotted as a function of FS angle.